\documentclass[a4paper,preprint,aps]{revtex4}

\usepackage{graphicx}
\usepackage{dcolumn}
\usepackage{bm}

\newcommand{\eqa}{\begin{equation}}
\newcommand{\eqz}{\end{equation}}
\newcommand{\eqma}{\begin{eqnarray}}
\newcommand{\eqmz}{\end{eqnarray}}

\begin{document}
\title{ Anharmonic force fields of perchloric acid, HClO$_4$, and perchloric anhydride, Cl$_2$O$_7$. An extreme case of inner polarization\thanks{In honor of Dr. Jean Demaison on his 60th birthday.}}
\author{A. Daniel Boese}
\affiliation{Institute of Nanotechnology, Forschungszentrum
Karlsruhe, P.O. Box 3640, D-76021 Karlsruhe, Germany}
\author{Jan M. L. Martin}
\email{comartin@wicc.weizmann.ac.il} \affiliation{Department of
Organic Chemistry, Weizmann Institute of Science, IL-76100
Re\d{h}ovot, Israel}
\date{{\em J. Mol. Struct.} (Jean Demaison issue) {\bf L05.38.05D}; Received May 26, 2005; Revised \today}
\smallskip
\begin{abstract}
DFT (density functional theory) anharmonic force fields with basis sets near the Kohn-Sham limit have been obtained for perchloric acid, HClO$_4$, and perchloric anhydride, Cl$_2$O$_7$. Calculated fundamental frequencies are in very good agreement with available experimental data. Some reassignments in the vibrational spectra of Cl$_2$O$_7$ are proposed based on our calculations. HClO$_4$ and Cl$_2$O$_7$ are particularly severe examples of the `inner polarization' phenomenon. The polarization consistent basis sets pc-1 and pc-2 (as well as their augmented counterparts) should be supplemented with two (preferably three) and one (preferably two) high-exponent $d$ functions, respectively, on second-row atoms. Complete anharmonic force fields are available as electronic supporting information\cite{ESI}.
\end{abstract}
\maketitle

\section{Introduction}

Perchloric acid, HClO$_4$, was first discovered nearly two centuries ago\cite{Stadion1818}. Its Raman spectra (and that of its anhydride Cl$_2$O$_7$) were first studied by Fonteyne\cite{Fonteyne} in the 1930s, who correctly concluded that the Cl$_2$O$_7$ molecule has a bridged structure but wrongly deduced C$_{3v}$ symmetry for HClO$_4$. Other early experimental spectroscopic work on these systems is reviewed by Karelin et al.\cite{Kar75} for HClO$_4$, and by Witt and Hammaker\cite{Wit73} for Cl$_2$O$_7$. 

Higher chlorine oxides have recently been implicated in theories of stratospheric destruction of ozone\cite{enviroshtuyot}. Despite the molecules' obvious importance, however, there have been rather few spectroscopic studies of them. The reason for the dearth of such studies is probably best illustrated by the following quote from Ref.\cite{Wit73}:
\begin{quotation}
{\em Caution!} Several explosions occurred during the course of this work. It was necessary to perform all experiments wearing heavy gloves and a face shield with the sample properly shielded.
\end{quotation}

Clearly, purely computational approaches do not entail such hazards.
In recent years, and following the pioneering study of Dressler and Thiel\cite{thiel}, DFT (density functional theory) has been considered by a number of groups as a cost-effective alternative for the calculation of molecular anharmonic force fields of medium-sized polyatomics. We cite, for instance, the work of Handy and coworkers on benzene\cite{anharmhandy2} as well as on furan, pyrrole and
thiophene\cite{Rudolf}, as well as simultaneous and independent studies by the group of Barone\cite{BaroneAzabenzenes} and the present authors\cite{anhar1} on the azabenzene series. Validation studies for a number of molecular anharmonic force fields were carried out simultaneously and independently by Barone and coworkers\cite{Baroneval,Baroneval2,Baroneval3,Baroneval4,Baroneval5} and by Boese, Klopper, and Martin\cite{anhar1,anhar2}.

As pointed out before\cite{Bau95,so2,so3,anhar2} --- both ab initio\cite{Bau95,so2} and DFT\cite{Bau95,so3,anhar2} --- basis set convergence in second-row compounds with a second-row atom in a high oxidation state presents a special challenge due to strong inner polarization effects.
(The effect has sometimes, e.g.,\cite{sio,PVX+dZ}, been 
referred to as `inner-shell polarization', but in 
fact can be shown to persist if the inner-shell orbitals are entirely replaced
by effective core potentials.\cite{cl2o7watoc}. Initially it was attributed to hypervalence\cite{Bau95}, but then shown\cite{sio} to occur in diatomic molecules that cannot plausibly be considered hypervalence. Alternative explanations that have been advanced include hypervalence\cite{Bau95}, polarization of the inner loops of the (3s,3p) valence orbitals\cite{PVX+dZ}, and improvement of the ability of the (3d) Rydberg orbital to accept backdonation from lone pairs of the surrounding atoms\cite{cl2o7watoc}.)
We will show that the subjects of the present study, perchloric acid (HClO$_4$) and perchloric anhydride (Cl$_2$O$_7$), are particularly extreme examples. We will also show that, even for a difficult `inorganic' molecule like Cl$_2$O$_7$, DFT anharmonic force fields can be quite useful in analyzing and assigning vibrational spectra.

The relationship between the present subject and the work of Dr. Jean Demaison, who is being honored by this special issue, is twofold. On the one hand, one of his research interests over the years has been high-resolution IR spectroscopy on second-row compounds, such as SiHF$_3$\cite{DemaisonSiHF3}, 
FPO\cite{DemaisonFPO}, FPS\cite{DemaisonFPS}, 
HNSO\cite{DemaisonHNSO}, SO$_2$F$_2$\cite{DemaisonSO2F2}, and many others, as well as ab initio and DFT electronic structure studies on such systems (e.g.,\cite{DemaisonSiF4,cf2}) and combined theoretical-experimental studies (e.g., on allyl phosphine\cite{DemaisonAllylphosphine}). On the other hand, Dr. Demaison and one of the present authors are fellow members in a IUPAC Task Group on the thermochemistry of radicals and other transient atmospheric and combustion species\cite{Volume1}. The title molecules are clearly relevant to this subject, and a somewhat accurate set of spectroscopic constants is indispensable for thermochemistry, particularly at elevated temperatures.

\section{Computational Details}

All calculations were run on the Linux farm (Intel Xeon and AMD Opteron)  of the Martin group  at the Weizmann Institute of Science.

Following the approach first proposed by Schneider and
Thiel\cite{rovib4}, a full cubic and semidiagonal quartic force field (all that is required for second-order rovibrational perturbation theory) are obtained by central numerical differentiation (in
rectilinear normal coordinates about the equilibrium geometry) of
analytical second derivatives. The latter were obtained by means of
locally modified versions of
{\sc gaussian~03}\cite{g03};
modified routines from {\sc cadpac}\cite{Cadpac} were used as the
driver for the numerical differentiation routine. 

All the force fields have been analyzed by means of the
{\sc spectro}\cite{Spectro} and by {\sc polyad}\cite{Polyad} rovibrational
perturbation theory programs developed by the Handy group and by Martin,
respectively.

A pruned (140$\times$974) quadrature grid was used, being the pruned direct product of 
a 140-point Euler-Maclaurin radial
grids\cite{Han93} and a 974-point Lebedev angular grid\cite{Lebedev}.
For the CPKS
(coupled-perturbed Kohn-Sham) steps, we used a different,
significantly coarser (75$\times$194) grid. It has been shown (e.g.,\cite{anhar1}) that this will affect computed fundamental frequencies by less than 1 cm$^{-1}$ while significantly reducing CPU time.

The numerical step size determined to be optimal in our previous
work\cite{anhar1} was:
\begin{eqnarray}
q_{step}(i)=\hbox{4}\times\sqrt{\frac{\mu}{\hbox{amu}}}\times
\sqrt{\frac{1000~\hbox{cm}^{-1}}{\omega(i)}}
\end{eqnarray}
The steps are done along the unnormalized Cartesian displacement
vector of the mass weighted normal coordinates.

We have furthermore tightened convergence criteria at all stages of
the electronic structure calculations to $10^{-10}$ or better (no
convergence could be achieved with even tighter criteria). 

The basis sets employed belong to the polarization consistent family of Jensen\cite{Jensen1,Jensen2,Jensen3,Jensen4}. Since bonding in the species investigated is highly ionic in character, (diffuse function) augmented basis sets\cite{Jensen3} were employed. We thus considered the aug-pc1, aug-pc2, and aug-pc3 basis sets, which are of $3s2p$, $4s3p2d$, and $6s5p3d2f$ quality, respectively, on hydrogen, of $4s3p2d$, $5s4p3d2f$, and $7s6p5d3f2g$ quality, respectively, on oxygen, and of $5s4p2d$, $6s5p3d2f$, and $7s6p5d3f2g$ quality, respectively, on chlorine. In addition, we considered aug-pc$m$+$n$d basis sets, in which $n$ high-exponent $d$ functions were added in an even-tempered series with stride factor 2.5 and starting from the highest-exponent $d$ function in the underlying basis set. 

Finally, the exchange-correlation functional employed is B97-1, which is the Handy group reparametrization\cite{HCTH93} of the Becke 1997 exchange-correlation functional. Our validation studies\cite{anhar1,anhar2} showed that, of the wide variety of functionals considered, B97-1 generally yields the best performance for harmonic and fundamental frequencies, marginally better than the very popular B3LYP functional.

\section{Results and discussion}

\subsection{Basis set convergence in binding energies, geometries and harmonic frequencies}

Geometries in Cartesian coordinates at all levels of theory can be downloaded as electronic supporting information to this paper\cite{ESI}. Our best geometries (B97-1/aug-pc3+d) are given in Figure 1.
Some salient bond distances are given in Table \ref{tab:Cl$_2$O$_7$dist}.

Inner polarization effects are particularly egregious for the central Cl--O--Cl bond distance in Cl$_2$O$_7$: 
we see a difference of no less than 0.083 \AA\ between aug-pc1 and our largest basis set (aug-pc3+d).  Adding a single tight $d$ function to Cl (resulting in the aug-pc1+d basis set) reduces the difference by 0.042 \AA. This is not nearly converged, however: the second and third tight $d$ function shave off another 0.014 and 0.004 \AA, respectively. The effect of the fourth tight $d$ is an order of magnitude smaller, indicating that convergence is reached. The difference between aug-pc1+4d and aug-pc3+d comes down to a not unreasonable 0.021 \AA. 

Expanding the underlying basis set to aug-pc2, we still find a contraction by 0.008 \AA\ upon adding the first $d$ function, and by an additional 0.002 \AA\ upon adding the second. The third tight $d$ function's effect is negligible. The difference with our largest basis set then stands at 0.0012 \AA, indicating satisfactory basis set convergence. (The aug-pc3 vs. aug-pc3+d differential is only a paltry 0.0002 \AA; consequently, no further basis set expansion was considered.)  

As noted repeatedly\cite{so2,so3,Wilson}, this type of effect on the geometry goes hand in hand with an inordinate basis set sensitivity of the computed total atomization energy, although Cl$_2$O$_7$ is, to the authors' knowledge, the most extreme example ever reported. Adding four tight $d$ functions to the aug-pc1 basis set increases the total atomization energy by no less than 100 (!) kcal/mol, individual contributions decreasing as 62:31:6:1. For the aug-pc2 basis set, the first added tight $d$ still has an effect of almost 15 kcal/mol, followed by 3.5 kcal/mol for the second and then decaying rapidly. The contribution of a tight $d$ to aug-pc3, 0.35 kcal/mol, is insignificant compared to the intrinsic error in DFT atomization energies. Overall basis set convergence in the aug-pc\{1,2,3\} sequence changes from a quite unsatisfactory \{302.2,411.0,429.7\} kcal/mol sequence without added tight $d$ functions to a rather more pleasing \{402.0,428.4,430.0\} kcal/mol.

Effects for the ClO distances in the ClO$_3$ groups are not much less severe: the aug-pc1 -- aug-pc2 -- aug-pc3 basis set increments of about 0.05 and 0.01 \AA, respectively, decrease by a factor of five in the aug-pc1+4d --- aug-pc2+3d --- aug-pc3+d sequence. 
  
Findings for the ClO$_3$ moiety in HClO$_4$ are similar, while those for the OCl distance in the HOCl moeity parallel the central OCl distances in Cl$_2$O$_7$ in their severity. As one might expect, the HO distance is not greatly affected.

As one could reasonably expect, inner polarization effects on the total atomization energy of HClO$_4$ are roughly half the size of the corresponding effects in Cl$_2$O$_7$. 

The largest effect in the harmonic frequencies will once more be seen in the ClO stretching modes.
The four such frequencies in HClO$_4$ are tabulated in Table \ref{tab:HClO$_4$freq}: a complete set of harmonic frequencies with all basis sets can once again be found as supporting information\cite{ESI}.

As one can see there, 1 cm$^{-1}$ convergence in harmonic frequencies requires at least three tight d functions for aug-pc1 (preferably four), and two for aug-pc2, while none are required for aug-pc3. Tight $d$ functions affect the aug-pc1 ClO stretches in the 90--130 cm$^{-1}$ range, compared to 8--20 cm$^{-1}$ for aug-pc2. We are not merely dealing with a systematic upward shift: for instance, the splitting between the two lower stretches in the ClO$_3$ moeity increases from 1 to 49 cm$^{-1}$ for the aug-pc1 basis set, and from 44 to 55 cm$^{-1}$ for aug-pc2, with concomitant shifts of intensity from the lower to the upper mode. The situation gets more problematic, if anything, for Cl$_2$O$_7$, where the lowest stretches interpenetrate with the upper bending frequencies, and the presence or absence of inner polarization functions indeed affects the ordering of bands.

HClO$_4$ and Cl$_2$O$_7$ thus far appear to represent the most extreme cases of inner polarization encountered in the literature. Clearly, pc-1 and aug-pc1 (and to a lesser extent, pc-2 and aug-pc2) basis sets should not be used in unmodified form for second-row compounds where a second-row atom finds itself in a high oxidation state. The pc-3 basis set, which already contains sufficiently high-exponent $d$ functions, appears to be immune to the problem.

\subsection{Anharmonic force field and comparison with earlier computational studies}

\subsubsection{Cl$_2$O$_7$}

A complete anharmonic force field using the aug-pc3+d basis set would be computationally intractable with available equipment: also, anharmonicities are fairly small. Therefore, we combined a B97-1/aug-pc1+2d quartic force field with the B97-1/aug-pc3+d geometry and harmonic frequencies. 

As expected, the anharmonicity constants involving the two low-lying torsion modes are physically unrealistic when determined by second-order rovibrational
perturbation theory: we have therefore decoupled those two modes from the remainder by zeroing all off-diagonal anharmonicity constants involving them.
A mild Fermi type 2 resonance $\nu_{16}+\nu_{18}\approx\nu_{3}$ was detected and taken into account.

Experimental work on the vibrational spectra of Cl$_2$O$_7$ is quite limited: the most complete experimental reference appears to be the work of Witt and Hammaker\cite{Wit73}. 

The only previous computational study on the vibrational spectrum of Cl$_2$O$_7$ that we are aware of is the work of Parthiban et al.\cite{parthiCl2O7}. These authors calculated HF/6-31G* harmonic frequencies and carried out a potential energy distribution analysis in terms of symmetry coordinates. 
Their calculated frequencies are in fair agreement with the observed IR spectrum of Witt and Hammaker\cite{Wit73}. Both sets of data are compared with our best computed harmonic and fundamental frequencies in Table \ref{tab:Cl$_2$O$_7$nu}.

Agreement between our computed fundamentals and experiment for the bands where the assigment is clear is too good to be merely coincidental: in general, we can expect agreement to within 20 cm$^{-1}$. Trying our hand at assigning the spectra summarized in Table 1 of Witt and Hammaker, we arrive at a considerably different assignment from those proposed by Witt and Hammaker (based on a simple empirical force field) and by Parthiban. A single glance should suffice to see that our (fairly high-level) force field has a very different structure from theirs. This illustrates the dangers of attempting to assign complex spectra based on low-level ab initio calculations or crude empirical force fields.

\subsubsection{HClO$_4$}

As this molecule is smaller to begin with, and anharmonicity effects are rather more important here (particularly for vibrations involving the H atom), we have calculated the anharmonic portion of the force field with a larger aug-pc2+2d basis set. Once again, the B97-1/aug-pc3+d geometry and harmonic frequencies were subsituted in the 2nd order rovibrational perturbation theory analysis. And similar to Cl$_2$O$_7$, the torsion mode was decoupled from the remaining ones in the analysis. No Fermi resonances sufficiently severe to require explicit treatment were detected. 

There is a high-resolution IR study\cite{Joh98} which reports $\nu_5$=726.99697(5) cm$^{-1}$ for H$^{35}$ClO$_4$ and 725.26209(9) cm$^{-1}$ for H$^{37}$ClO$_4$. (For comparison, our calculated fundamental is 711 cm$^{-1}$.) A number of lower-resolution studies have been reviewed and summarized by Karelin\cite{Kar75,Kar97}.

Francisco\cite{Fra95} found fairly good agreement between his computed MP2/6-31G(2d,2p) frequencies and the expt. data of Gigu\`ere and Savoie\cite{Gig62}.
Karelin\cite{Kar75,Kar97} noted that some of the bands assigned to HClO$_4$
by these latter two authors actually belong to Cl$_2$O$_7$ impurities. As shown in Table \ref{tab:HClO$_4$nu}, the Karelin data agree quite well with our calculated anharmonic frequencies, except of course for the low torsion band which was not observed experimentally. 

The large discrepancy between the frequencies of Francisco and ourselves for the low torsion mode puzzled us. Upon recalculation at the same level of theory as used by Francisco, MP2/6-31G(2d,2p), we found that we are unable to reproduce his data. The recalculated frequencies agree as well as expected with our large basis set DFT data.

Oberhammer and coworkers\cite{Cas94} obtained a computed geometry from a combination of gas electron diffraction, microwave spectroscopy, and (to fix the difference between the two unique ClO distances in the ClO$_3$ group) ab initio calculations. Our geometry agrees (Table \ref{ref:HClO$_4$rg}) as well with theirs as can be expected considering the limitations of the respective computational and experimental techniques. For the benefit of future experimental workers, we
have included anharmonic zero-point corrections to the geometry in Table \ref{ref:HClO$_4$rg}.

\section{Conclusions}

We may draw the following conclusions:
\begin{itemize}
\item DFT anharmonic force fields are a relatively cost-effective way to assign vibrational spectra of medium-sized polyatomics
\item some reassignments of the vibrational spectrum of Cl$_2$O$_7$ may be in order
\item HClO$_4$ and Cl$_2$O$_7$ are particularly severe examples of the `inner polarization' phenomenon. Consideration of previous results\cite{so3,anhar2} as well as unpublished data by our group on SF$_6$ and PF$_5$, suggests that there is a positive correlation between the oxidation state of the central second-row atom atom in a molecule (or group) and the severity of the phenomenon
\item the pc1 and pc2 basis sets (as well as their augmented counterparts) should not be used in unmodified form for systems liable to be affected by this phenomenon. We recommend the addition of two (preferably three) high-exponent d functions to second-row atoms in the pc1 basis set, and one (preferably two) such function(s) for the pc2 basis set. 
\end{itemize}
It is often assumed that basis set convergence in DFT is not a very
important issue, and that "push-button" standard basis sets suffice. While the
absence of an explicit interelectronic cusp does mean DFT basis set
convergence is a much less acute problem than in correlated ab initio methods (and in fact
fairly similar to Hartree-Fock theory in that regard), considerable
basis set dependence remains\cite{Mar2000,BMH2003}. And in fact, any serious
basis set convergence issue that shows up at the Hartree-Fock level --- such as inner polarization\cite{so2} ---
will also be seen in DFT. Cl$_2$O$_7$ and HClO$_4$ are just
particularly dramatic
examples thereof.

\section{Acknowledgments}
ADB acknowledges a postdoctoral fellowship from the Feinberg
Graduate School (Weizmann Institute). Research at Weizmann was
supported by the Minerva Foundation, Munich, Germany, by the Lise
Meitner-Minerva Center for Computational Quantum Chemistry (of which
JMLM is a member {\em ad personam}), and by the Helen and Martin Kimmel Center for
Molecular Design. This work is related to Project 2003-024-1-100, "Selected Free Radicals and Critical Intermediates: Thermodynamic Properties from Theory and Experiment," of the International Union of Pure and Applied Chemistry (IUPAC).

 \indent

\clearpage

\begin{table}
\caption{Basis set convergence for bond distances (\AA) and total atomization energies (kcal/mol)
in Cl$_2$O$_7$ and HClO$_4$. The B97-1 functional was used throughout.\label{tab:Cl$_2$O$_7$dist}}
\begin{tabular}{lccccc}
\hline\hline
HClO$_4$   & r(Cl-O) &r(Cl=O)  &r(Cl=O)   &r(Cl-H)  &     TAE  \\
           &         &trans    &gauche\\
           \hline
aug-pc1    & 1.7331  & 1.4664  & 1.4764   &0.9778  &     266.26  \\
aug-pc1+d  & 1.6878  & 1.4279  & 1.4377   &0.9769  &     297.93  \\
aug-pc1+2d & 1.6731  & 1.4167  & 1.4263   &0.9764  &     313.69  \\
aug-pc1+3d & 1.6693  & 1.4139  & 1.4235   &0.9764  &     316.77  \\
aug-pc1+4d & 1.6689  & 1.4137  & 1.4232   &0.9763  &     317.26  \\
aug-pc2    & 1.6571  & 1.4152  & 1.4247   &0.9710  &     324.65  \\
aug-pc2+d  & 1.6501  & 1.4104  & 1.4197   &0.9709  &     331.64  \\
aug-pc2+2d & 1.6484  & 1.4090  & 1.4184   &0.9708  &     333.44  \\
aug-pc2+3d & 1.6483  & 1.4090  & 1.4183   &0.9708  &     333.54  \\
aug-pc3    & 1.6471  & 1.4069  & 1.4162   &0.9704  &     334.22  \\
aug-pc3+d  & 1.6469  & 1.4068  & 1.4161   &0.9704  &     334.40  \\
\hline
Cl$_2$O$_7$& r(Cl-O) & r(Cl=O) & r(Cl=O)  &r(Cl=O) &     TAE  \\
           & bridge  & trans   & gauche1  &gauche2\\
\hline
aug-pc1    & 1.8087  & 1.4693  & 1.4713   &1.4694  &     302.33  \\
aug-pc1+d  & 1.7662  & 1.4298  & 1.4325   &1.4307  &     364.15  \\
aug-pc1+2d & 1.7520  & 1.4182  & 1.4209   &1.4193  &     394.99  \\
aug-pc1+3d & 1.7479  & 1.4153  & 1.4181   &1.4165  &     401.02  \\
aug-pc1+4d & 1.7476  & 1.4150  & 1.4178   &1.4162  &     401.98  \\
aug-pc2    & 1.7365  & 1.4160  & 1.4188   &1.4173  &     410.98  \\
aug-pc2+d  & 1.7290  & 1.4109  & 1.4138   &1.4123  &     424.66  \\
aug-pc2+2d & 1.7272  & 1.4095  & 1.4124   &1.4109  &     428.19  \\
aug-pc2+3d & 1.7271  & 1.4094  & 1.4123   &1.4108  &     428.39  \\
aug-pc3    & 1.7262  & 1.4073  & 1.4102   &1.4087  &     429.68  \\
aug-pc3+d  & 1.7260  & 1.4071  & 1.4101   &1.4086  &     430.03  \\
\hline\hline
\end{tabular}
\end{table}

\begin{table}
\caption{Basis set convergence for the four ClO stretching frequencies (cm$^{-1}$, intensities in km.mol$^{-1}$ in parentheses) of HClO$_4$. The B97-1 functional was used throughout.\label{tab:HClO$_4$freq}}
\begin{tabular}{ll}
\hline\hline
aug-pc1   & 927.5($a'$,77.35);1113.6($a'$,139.8);1114.7($a''$,217.1);1234.8($a'$,128.9)\\
aug-pc1+d & 998.4($a'$,80.33);1177.7($a'$,116.8);1202.3($a''$,256.2);1287.5($a'$,192.2)\\
aug-pc1+2d&1028.3($a'$,79.54);1195.7($a'$,99.40);1240.6($a''$,269.2);1313.2($a'$,224.3)\\
aug-pc1+3d&1034.0($a'$,79.75);1199.0($a'$,98.09);1246.7($a''$,272.5);1318.3($a'$,229.1)\\
aug-pc1+4d&1035.0($a'$,79.68);1199.4($a'$,97.61);1248.0($a''$,272.9);1319.2($a'$,230.0)\\
aug-pc2   &1038.7($a'$,82.66);1205.9($a'$,101.6);1249.8($a''$,263.6);1323.5($a'$,215.6)\\
aug-pc2+d &1051.8($a'$,82.93);1212.1($a'$,96.77);1265.0($a''$,270.2);1334.9($a'$,227.8)\\
aug-pc2+2d&1055.0($a'$,82.73);1213.5($a'$,95.70);1268.6($a''$,271.6);1337.8($a'$,230.4)\\
aug-pc2+3d&1055.2($a'$,82.74);1213.6($a'$,95.64);1268.9($a''$,271.7);1338.0($a'$,230.6)\\
aug-pc3   &1057.3($a'$,81.77);1213.9($a'$,95.64);1269.6($a''$,272.3);1338.6($a'$,231.5)\\
aug-pc3+d &1057.6($a'$,81.77);1214.1($a'$,95.56);1269.9($a''$,272.4);1338.9($a'$,231.7)\\
\hline\hline
\end{tabular}
\end{table}

\begin{table}
\caption{Computed and observed harmonic and fundamental frequencies (cm$^{-1}$) for Cl$_2$O$_7$\label{tab:Cl$_2$O$_7$nu}. Infrared intensities in km/mol given 
in   parentheses with harmonic frequencies.}
\begin{tabular}{lccccl}
\hline\hline
\multicolumn{3}{c}{Parthiban et al.\cite{parthiCl2O7}}&\multicolumn{3}{c}{Present work}\\
(*)&HF/6-31G*&assignment&best $\omega_i$&best $\nu_i$& assignment\\
\hline
\multicolumn{6}{c}{A symmetry block}\\
S1   &  1383.9  & 1300 & 1324(320)  &1297   & 1313a \\
S2   &  1341.7  & 1300 & 1292(2)    &1265   & 1260a \\
S3   &  1129.4  & 1060 & 1076(13)   &1054b  & 1057 \\
S4   &   819.6  &  704 &  715(39)   & 706   & 704 \\
S5   &   553.4  &  567 &  642(3)    & 633   & 639(s) c \\
S6   &   604.9  &  600 &  562(0.02) & 553   & -- \\
S7   &   715.6  &  512 &  512(7)    & 503   & 521 d \\
S8   &   324.6  &  283 &  291(0.008)& 287   & $\backslash$ 283(g),295(l),286(l),272(l) \\
S9   &   316.0  &  283 &  282(0.02) & 278   & $/$ \\
S10  &   167.4  &  154 &  145(0.2)  & 138   & 154 \\
S11  &    53.7  &   -- &   30(0.001)&  --   & -- \\
\multicolumn{6}{c}{B symmetry block}\\
S12  &  1371.4  & 1300 & 1315(386)  &1289   &1300a \\
S13  &  1352.3  & 1300 & 1301(71)   &1274   &1274a \\
S14  &  1078.9  & 1025 & 1039(146)  &1020   &1029R,1025Q,1020P \\
S15  &   752.8  &  698 &  592(12)   & 582   & 600 \\
S16  &   582.5  &  571 &  578(103)  & 571   & 571(g),565(s) \\
S17  &   636.6  &  600 &  571(268)  & 564   & 555(s) \\
S18  &   629.5  &  488?&  508(305)  & 498   & 512 d \\
S19  &   466.0  &  488 &  429(11)   & 421   & 430 e \\
S20  &   304.2  &  272 &  270(39)   & 267   & 272 \\
S21  &    86.9  &   -- &   87(0.04) &  --   & -- \\
\hline\hline
\end{tabular}

(*) symmetry coordinate most represented in normal mode. For definitions, see Ref.\cite{parthiCl2O7}\\
(a) 1300 cm$^{-1}$ band resolves to 4 bands in Ar matrix at 20 K;
see footnote b of Table 1 of Ref.\cite{Wit73}\\
(b) Mild Fermi 2 resonance with $\nu_{16}+\nu_{18}=$1069 cm$^{-1}$\\
(c) assigned to Cl$_2$O impurity by Witt and Hammaker. Gas-phase high-resolution data for Cl$_2$O: $\nu_1$=641.9694(1) cm$^{-1}$ and $\nu_3$=686.5936(1) cm$^{-1}$\cite{XuCl2O}\\
(d) Witt and Hammaker propose opposite assignment on force field grounds\\
(e) assigned to overtone 272+154 by Witt and Hammaker

\end{table}

\begin{table}
\caption{Computed and observed harmonic and fundamental frequencies (cm$^{-1}$, IR intensities in km/mol in parentheses) for HClO$_4$\label{tab:HClO$_4$nu}}
\begin{tabular}{cccccc}
\hline\hline
\multicolumn{2}{c}{MP2/6-31G(2d,2p)}&Expt. & Expt. & \multicolumn{2}{c}{This work}\\
Ref.\cite{Fra95}&this work&Ref.\cite{Gig62}&Ref.\cite{Kar97}&best $\omega_i$& best $\nu_i$\\
$a'$\\
3554  & 3748&3560& 3553 &   3743.8(119.8) & 3557\\
1225  & 1361&1263& 1326 &   1338.9(231.7) & 1307\\
1215  & 1295&1200& 1201 &   1214.1(95.56) & 1180\\
1016  & 1069&1050& 1048 &   1057.6(81.77) & 1038\\
 690  &  704& 725&  726 &    725.4(185.6) &  711\\
 551  &  555& 560&  582 &    576.4(17.64) &  569\\
 526  &  539& 519&  555 &    557.4(4.053) &  548\\
 395  &  395& 390&  421 &    412.0(4.750) &  405\\
$a''$\\
1333  & 1295&1326& 1265 &   1269.9(272.4) & 1242\\
 552  &  558& 579&  582 &    583.2(22.41) &  576\\
 500  &  403& 430&  421 &    420.5(11.41) &  408\\
 357  &  193& 307&  --- &    190.8(84.72) &  191\\
\hline\hline
\end{tabular}
\end{table}

\begin{table}
\caption{Computed and observed geometry of HClO$_4$\label{ref:HClO$_4$rg}}
\begin{tabular}{cccc}
\hline\hline
Ref.\cite{Cas94}   & \multicolumn{3}{c}{B97-1/aug-pc3+d}\\
GED/MW   & $r_e$   &  $r_g-r_e$ &  $r_z-r_e$   \\
\hline
1.404(1) & 1.4068 &  0.0070 &  0.0056   \\
1.414(1) & 1.4161 &  0.0065 &  0.0051  \\
1.641(2) & 1.6469 &  0.0106 &  0.0095  \\
(0.98)   & 0.9704 &  0.0303 & -0.0219   \\
115.0(2) & 115.00\\
114.6(2) & 113.61\\
101.5(15)& 100.93\\
104.2(8) & 105.17\\
(105.0)  & 105.66\\
\hline\hline
\end{tabular}
\end{table}

\newpage
\begin{figure}
\includegraphics{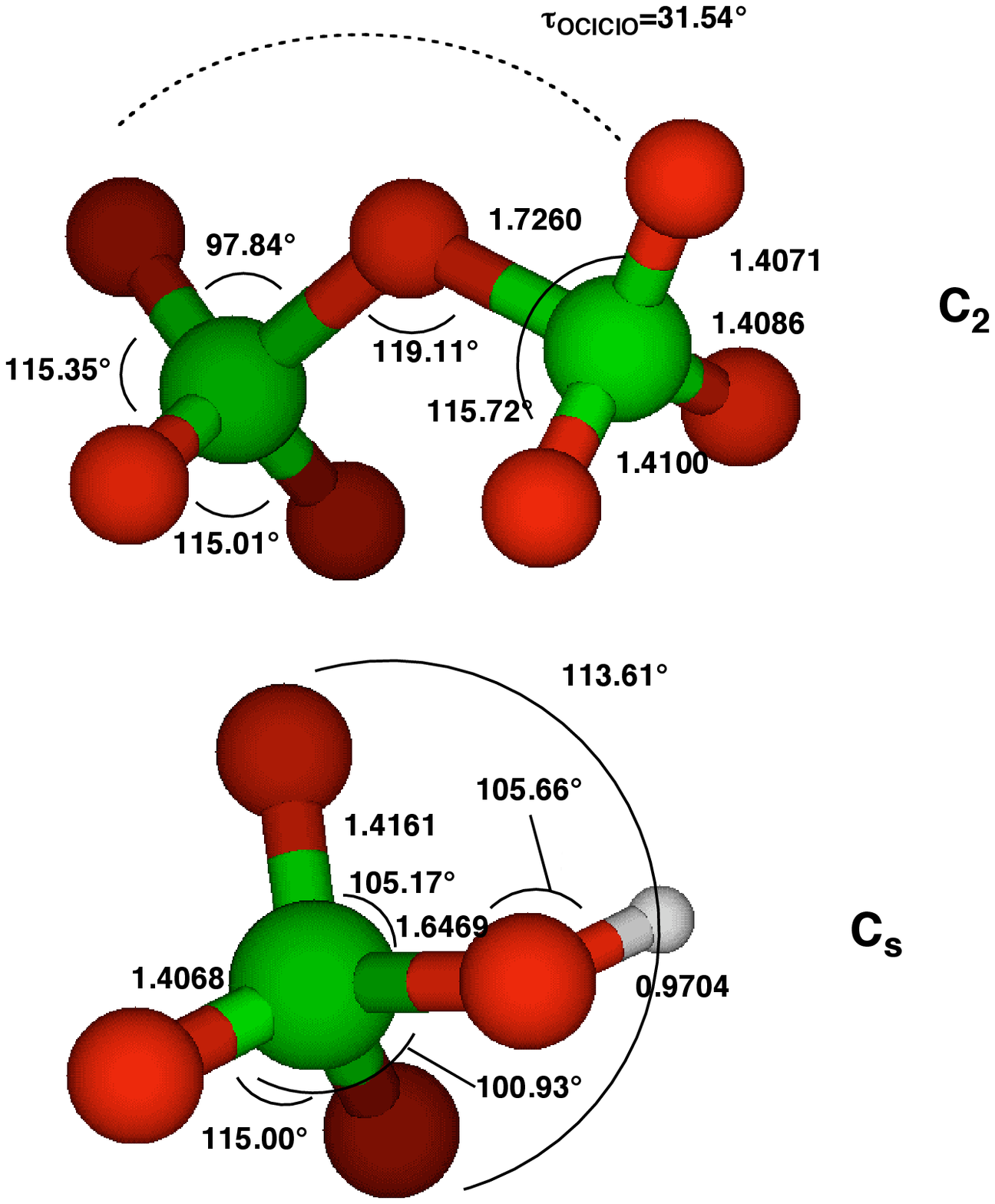} \\
\caption{B97-1/aug-pc3+d geometries (\AA, degrees)}
\end{figure}

\end{document}